# Comments on "Phonon and magnetic structure in δ-plutonium from density-functional theory" by P. Söderlind *et al.*

M. Janoschek[1], G. H. Lander[2], J. M. Lawrence[1], E. D. Bauer[1], M. D. Lumsden[3], D. L. Abernathy[3], and J. D. Thompson[1]

[1]*Los Alamos National Laboratory (LANL), Los Alamos, NM 87545, USA.*
[2]*European Commission, Joint Research Centre, Institute for Transuranium Elements, D-76125 Karlsruhe, Germany.*
[3]*Quantum Condensed Matter Division, Oak Ridge National Laboratory, Oak Ridge, TN 37831-6475, USA.*

**Abstract**
In their recent paper [Söderlind, P. *et al.*, Sci. Rep. 5, 15958 (2015)], Söderlind *et al.* discuss two subjects from a theoretical point of view: the phonon spectra and the possible magnetic structure of δ-plutonium (Pu). Here, we comment on the second subject. Söderlind *et al.* compare the Pu magnetic form factor $F(Q)$ calculated via density functional theory (DFT) with measurements of $F(Q)$ by neutron spectroscopy [Janoschek, M. *et al.*, Sci. Adv. 1, e1500188 (2015)]. In particular, this comparison does not consider a number of experimental facts established in the neutron spectroscopy study.

## 1. Introduction
In Ref. [1] Söderlind *et al.* present further developments of their DFT code and corresponding results obtained on Pu metal. As summarized by the authors, due to its remarkable ambient pressure phase diagram with six allotropic phases below its melting temperature and with large volumetric changes of up to 25% between these phases, Pu represents a formidable challenge to solid state theory. On the basis of a picture in which the *5f* electrons of Pu are itinerant, the authors of Ref. [1] successfully use their DFT framework to explain the main features of Pu's phase diagram [3]. They also demonstrate that DFT can be used to calculate phonon spectra for the δ-phase of Pu and that the calculations are in reasonable agreement with the spectra determined via inelastic x-ray scattering [4].

In addition, Ref. [1] suggests a solution to Pu's magnetism. The authors of Ref. [1] have a history of addressing the issue of magnetism in δ-Pu on the basis of DFT calculations dating back to at least 2004 [3]. At that time, they proposed long-range *magnetic order* in plutonium, including its α- and δ-Pu phases. (In this comment, we discuss only δ-Pu, as this phase is the only one examined in Ref. [2]). In 2005 Lashley *et al.* [5] surveyed the experimental situation and stated that there was no experimental evidence for long-range *static* magnetic order in δ-Pu. In addition, the absence of magnetic scattering intensity between 0 and 20 meV unambiguously demonstrated that *quasi-elastic* fluctuations of magnetic character also could be excluded [5]. In contrast, *inelastic* magnetic fluctuations for energies larger than 20 meV could not be addressed because no experiments were available at that time. Apart from neutron-scattering experiments, the absence of magnetism in Pu was further confirmed by a number of experiments, most notably muon spin-rotation experiments [6] that *excluded the possibility of any static long-range magnetic order or slow magnetic fluctuations with magnetic moments greater than $10^{-3}$ $\mu_B$.*

Reference [1] proposes to reconcile DFT theory with the absence of magnetic signatures in experiments on δ-Pu by introducing *disordered local* magnetic moments that exhibit a special character wherein the spin ($\mu_S$) and orbital ($\mu_L$) moments have *exactly* the same magnitude but are opposite in sign so that the net moment is *zero*. Because the orbital and spin moments have a different spatial extent, however, the authors suggest that the experimental magnetic form factor reported in Ref. [2] arises from this effect – the cancellation of the spin and orbital contributions. In the following, we summarize some critical experimental facts established in our neutron-scattering work [2] and compare them to the model of disordered local magnetic moments in Ref. [1].

## 2. Relevance of the error bars associated with our neutron spectroscopy data
The authors of Ref. [1] are, of course, correct that a cancellation of $\mu_L$ and $\mu_S$ would result in a magnetic form factor $F(Q)$ resembling the $\langle j_2 \rangle$ Bessel function (see Fig. 1). In particular, for that case $F(Q)$ vanishes for momentum transfers Q → 0. It is further true that the measured magnetic form factor decreases as Q → 0; however, the data do not support the conclusion that $F(Q→ 0) → 0$, as we demonstrate in the following. In Fig. 1 we plot the experimentally determined $F(Q)$ for δ-Pu obtained in two different measurements performed at two incident neutron energies $E_i$ = 250 and 500 meV (black squares and empty circles in Fig. 1). We compare these data to the results of DFT calculations presented in Ref. [1] (green line) and to results from Dynamical Mean Field Theory (DMFT) calculations (blue line) [2]. In their comparison to data, Söderlind et al. neglect the statistical error bars on data points and argue that most of them are small [1]. As shown in Fig. 1, however, the statistical errors in the limit $Q→ 0$ are particularly large, and, further, the experiment does not provide data points for Q/4π < 0.12 Å$^{-1}$. The authors [1] further claim that the "DMFT model agrees almost as well with the neutron-spectroscopy data [as the DFT] but seems to lack the correct functional form for small Q." A comparison based solely on the values of $F(Q)$ may suggest better agreement between DFT calculations and the data than between DMFT calculations and data, but taking into account experimental error bars paints an entirely different picture. To demonstrate this, we define in the standard way an agreement factor

$$R = \frac{\sum_i w_i [F^{exp}(Q_i) - F^{calc}(Q_i)]^2}{\sum_i w_i [F^{exp}(Q_i)]^2},$$

where $F^{exp}(Q_i)$ and $F^{calc}(Q_i)$ are the experimentally determined and calculated magnetic form factors at momentum transfer $Q_i$, and $w_i$ is the weight of each data point given by the corresponding error bar. In calculating the agreement factor, we combine both data sets. We find $R^{DFT}$ = 0.40 and $R^{DMFT}$ = 0.33 for the DFT and DMFT calculations, respectively. This straightforward calculation demonstrates that the DMFT results agree somewhat better with experiment, in contrast to the remarks in Ref. [1]. Apart from the large error bars for Q/4π < 0.2 Å$^{-1}$, the discrepancy between conclusions when error bars are excluded or appropriately taken into account arises because DFT produces a much narrower peak that does not agree with data in the $Q$-range between 0.2 Å$^{-1}$ < Q/4π < 0.3 Å$^{-1}$ where the error bars are small.

It is also instructive to discuss why the error bars for $Q/4π$ < 0.2 Å$^{-1}$ are large. As we will discuss in more detail below, the experimental magnetic form factor has been determined from inelastic neutron-scattering measurements from a magnetic signal with a spin-fluctuation energy of 84 meV. Because of the finite mass of the neutron, scattering processes at large energy transfer and small $Q$ are restricted by kinematics. This is easily visible by inspecting the full energy and momentum-transfer dependence of the magnetic signal that is shown in Fig. 1 (E) in Ref. [2]. These are well-known effects in neutron scattering, and for the used incident energies, in particular for the neutron-energy transfers at which this magnetic resonance is observed, quantitative analysis of the scattering intensity below Q/4π = 0.2 Å$^{-1}$ has to be regarded with caution. We note that the error bars plotted in Fig. 1 are only the statistical error bars stemming from uncertainty in counting neutrons; however, more representative error bars that include uncertainty due the restricted neutron kinematics are likely to be even larger for small momentum transfers. This is explained in detail in the supplementary information of Ref. [2].

The Q dependence of the magnetic form factor for small momentum transfers has to be determined via *elastic* neutron scattering experiments on single crystals in high magnetic fields, as for example carried out on PuSb [7]. We note that currently there are no large enough single crystals of δ-Pu to allow such an experiment. To conclude, the currently available data from Ref. [2] do not allow a a definitive statement that $F(Q→ 0) → 0$, but the data that are most reliable (have the smallest error bars) suggest that this is not the case.

## 3. Absence of quasielastic scattering
The possibility of disordered local magnetic moments, such as those proposed by Söderlind et al. for δ-Pu [1], have been examined extensively by neutron-scattering experiments. Entropy considerations usually lead to magnetic ordering of such disordered moments at low temperatures, but in some cases, for example in frustrated systems or spin glasses, ordering does not occur. As demonstrated early on in the *first* study of antiferromagnetism in 1949 [8], such disordered moments can be detected by means of neutron spectroscopy via their fluctuations above the ordering temperature. For materials without strong uniaxial magnetic anisotropy, the dynamic response of such fluctuating disordered moments observed by neutron spectroscopy is a *quasielastic* signal centered on zero neutron-energy transfer with an energy width (FWHM) that *is always of the order of thermal energies $k_B T$*. The first neutron-spectroscopy experiment on δ-Pu was carried out originally to investigate the phonon density of states [9]. In a later paper [5] these data were reanalyzed to verify if, in the energy window available in those experiments (± 20 meV), any such quasielastic magnetic scattering due to fluctuating disordered moments were present. Because the experiment was carried out for temperatures at and below room temperature, the energy width of the corresponding quasielastic scattering should be equal to or less than 25 meV. Because no magnetic contribution was found in this energy window, the existence of fluctuating disordered magnetic moments can be excluded. For completeness, we note that, although the specific model of disordered local moments proposed in Ref. [1] for Pu considers the magnitude of the fluctuating magnetic moment to be zero (due to the cancellation of spin and orbital components), the corresponding fluctuations should be still observed via neutron scattering. This is because the corresponding magnetic form factor $F(Q)$ is nonzero for $Q \neq 0$ (cf. Fig. 1). The $Q$-range analyzed by Lashley *et al.* [5] corresponds to 0.06 Å$^{-1}$ < $Q/4\pi$ < 0.16 Å$^{-1}$ for which $F(Q)$ approaches ≈ 0.5.

## 4. Inelastic nature of the magnetic scattering

Reference [1] suggests that our neutron-spectroscopy results on δ-Pu [2] contradict the earlier analysis by Lashley *et al.* [5,9]. In stark contrast, our findings support those findings that no *quasielastic* magnetic fluctuations exist in δ-Pu due disordered magnetic moments. Notably, our experiments identify an *inelastic feature,* namely a *spin resonance* that is characterized by a spin-fluctuation energy $E_{sf}$ = 84 meV, instead of *quasielastic* spin fluctuations centered at zero energy associated with fluctuations of disordered local moments (see discussion in (2)). We note that the *inelastic* nature of the spin fluctuations is not discussed in Ref. [1]. Our results [2], as well as the earlier neutron-spectroscopy studies [5, 9], are supported by muon spin-rotation experiments [6] that come to the same conclusion, and we can conclude safely that there are no fluctuating disordered local moments in δ-Pu.

Reference [1] proposes that the reason for the absence of long-time average (*static*) magnetic moments is the cancellation of orbital and spin moments. In contrast, the presence of *a spin resonance* at $E_{sf}$ = 84 meV [2] implies that this is because the Pu magnetic moment is screened via the spins of conduction electrons and no long-time average exists. In combination with results from core-hole photoemission spectroscopy (CHPES) [10] and resonant x-ray emission spectroscopy (RXES) [11] that probe the valence configuration, neutron spectroscopy shows that the screening of static magnetic moments is driven by virtual valence fluctuations. The details of this mechanism are explained in Ref. [2] and follow the well-established reasoning for valence-fluctuating compounds, such as α-Ce [12], CePd$_3$ [13] and others that are discussed in Ref. [14].

A spin resonance, such as that observed in Ref. [2], was predicted from DMFT calculations in 2007 (before the experiment) [15], and the spin-fluctuation energy was calculated to be $E_{sf}$ ≈ 70 meV. Other theories of this type gave values between 63 and 187 meV [16, 17]. The results of Ref. [2] are a confirmation of this type of theory. We note that DFT (as used in Ref. [1]) is a ground-state theory and as such cannot predict an excited state property like the observed spin resonance at $E_{sf}$ = 84 meV.

## 5. Size of the magnetic moment

As discussed above, the absence of magnetic scattering at and around zero energy [5, 9] demonstrates the absence of *static* or *quasielastic* magnetic moments, respectively. By energy-integrating the spin resonance observed in Ref. [2], however, it can be shown that a non-zero dynamic magnetic moment exists with a lifetime (τ ≈ 0.02 ps) because of coupled spin and valence fluctuations. The size of this magnetic moment is 0.6(2) μ$_B$. This observation is incompatible with the proposition in Ref. [1] that spin and orbital components of the magnetic moment cancel each other. In particular, the statement in Ref. [1] that the DFT result of "magnetic disorder simply represents a frozen (static) state of the fluctuations (observed in our work [2])" is not correct. As discussed above in detail, such static or quasielastic fluctuations do not exist in δ-Pu. Notably, the magnetic moment discussed in Ref. [2] is an excited state property and ceases to exist after the lifetime of the fluctuation, and thus the static magnetic moment is zero.

## 6. Itinerant vs localized f electrons
Reference [1] states that "the DMFT [2] models the 5*f*-electron states as superposition of localized 5*f* wave functions with screened magnetic moments while DFT describes the 5*f* electrons as itinerant with spin and orbital moments effectively cancelling each other." This statement about DMFT is not correct. As explained in Ref. [2], DMFT considers hybridization of localized 5*f* electrons with conduction electrons. This implies that the 5*f* electrons in DMFT models are neither fully itinerant nor localized. It is well-accepted in the literature on Pu that its 5*f*-electron configuration in the ground state is neither perfectly itinerant nor localized [18, 19].

## 7. Comparison to *F*(*Q*) of Sm in SmCo$_5$
A comparison to *F*(Q) of SmCo$_5$, as proposed in Ref. [1], is not justified. The case of Sm$^{3+}$ is well understood in terms of free-ion form factors [20]. For Sm$^{3+}$ there is significant mixing at room temperature of the total angular momentum $J = 7/2$ excited state into the $J = 5/2$ ground state, as the energy difference between these states is ~ 120 meV. Such mixing does not occur in Pu materials because the spin-orbit interaction is greater in Pu than in Sm, and the splitting is ~ 400 meV (see supplementary information of Ref. [2]).

## 8. Conclusions
In summary, when error bars on inelastic neutron-scattering data are considered correctly, DMFT calculations reproduces the measured magnetic form factor somewhat better than DFT calculations presented in Ref. [1]. Parenthetically, we note that error bars on the phonon-dispersion of δ-Pu measured by Wong *et al.* [4] also have been omitted in a comparison of DFT and DMFT results [1] and should be included for meaningful statements. In addition, we show that the results of previous [5,9] and our more recent neutron spectroscopy results [2], as well as the results of muon spin-rotation experiments [6] on δ-Pu, are not compatible with the model of disordered local moments proposed in Ref. [1]. The model in Ref. [1] is unable to explain that the magnetic signature of δ-Pu is a spin resonance with a spin-fluctuation energy $E_{sf}$ = 84 meV and fails to reproduce the observed dynamic magnetic moment of 0.6(2) $\mu_B$. In contrast, DMFT [2, 7] correctly accounts for the spin-fluctuation energy, the size of the dynamic magnetic moment, the magnetic form factor, and the lifetime of associated valence fluctuations as well as the occupation of the three valence configurations that form the valence-fluctuating electronic ground state of δ-Pu. In addition, the valence-fluctuating ground state proposed by Shim *et al.* [7] also provides a natural explanation of the observed static magnetic susceptibility that is nearly temperature-independent. Using the sum-rule for the dynamic magnetic susceptibility from neutron spectroscopy, [2] we have shown that this model can account quantitatively for the static magnetic susceptibility measured in Ref. [21]. Thus, DMFT [2,7] provides a rather complete description of the missing static magnetism in δ-Pu.

### Acknowledgments
We acknowledge useful discussions with Jian-Xin Zhu.

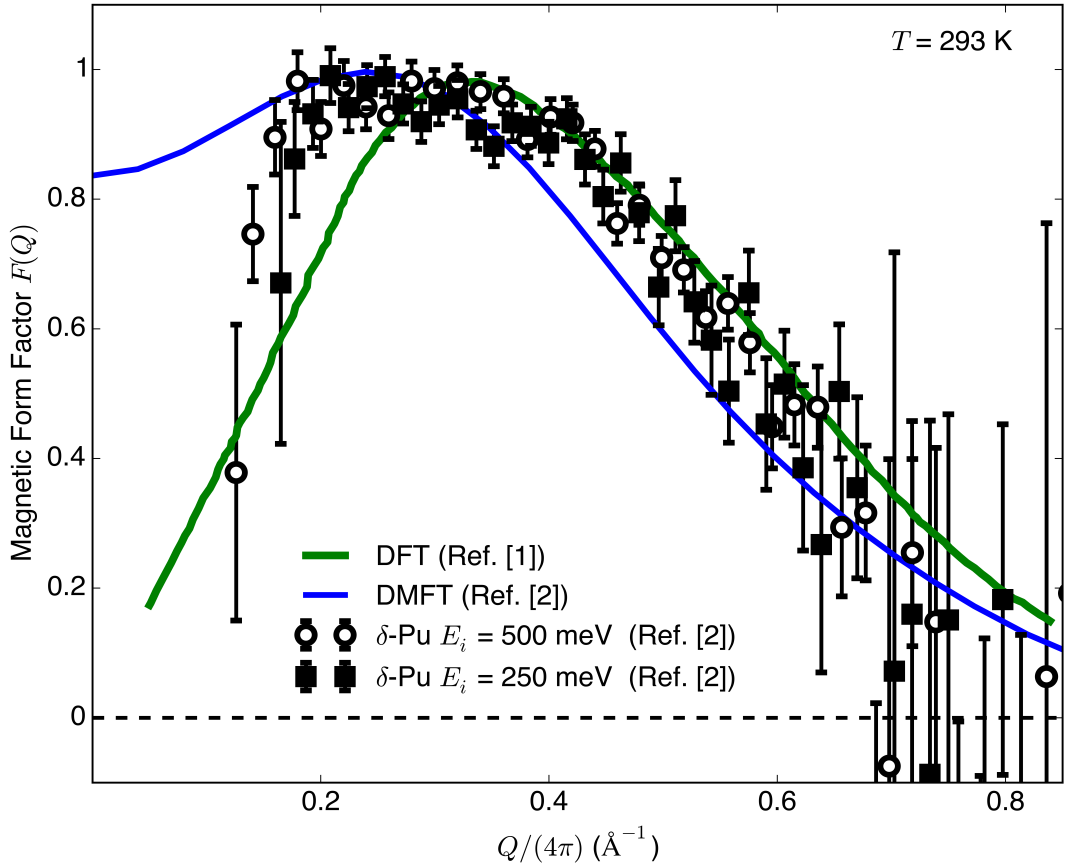

**Fig. 1. The magnetic form-factor for δ-plutonium (δ-Pu).** The black squares and open circles denote the magnetic form factor for δ-Pu as determined by our neutron spectroscopy experiment carried out at room temperature (T = 293 K) and with incident neutron energies $E_i$ = 250 and 500 meV, respectively [2]. The solid green and blue lines are the results of the DFT calculations [1] and of Dynamical Mean Field Theory (DMFT) calculations [2], respectively.